\documentstyle[12pt]{article}
\begin{document}

\title{\textbf{Generalized second law of thermodynamics in modified FRW
cosmology with corrected entropy-area relation}}

\author{K. Karami$^{1,2}$\thanks{E-mail: KKarami@uok.ac.ir} , A. Sheykhi$^{3,2}$\thanks{E-mail: sheykhi@mail.uk.ac.ir} ,
N. Sahraei$^{1}$\thanks{E-mail: nedasahraei@ymail.com} , S. Ghaffari$^{1}$\thanks{E-mail: ghaffari$_{-}$sh113@yahoo.com}\\\\
$^{1}$\small{Department of Physics, University of Kurdistan,
Pasdaran St., Sanandaj, Iran}\\$^{2}$\small{Research Institute for
Astronomy $\&$ Astrophysics of Maragha (RIAAM), Maragha, Iran}\\
$^{3}$\small{Department of Physics, Shahid Bahonar University, P.O.
Box 76175, Kerman, Iran}}

\maketitle

\begin{abstract}
Using the corrected entropy-area relation motivated by the loop
quantum gravity, we investigate the validity of the generalized
second law of thermodynamics in the framework of modified FRW
cosmology. We consider a non-flat universe filled with an
interacting viscous dark energy with dark matter and radiation.
The boundary of the universe is assumed to be the dynamical
apparent horizon. We find out that the generalized second law is
always satisfied throughout the history of the universe for any
spatial curvature regardless of the dark energy model.
\end{abstract}


\newpage
\section{Introduction}
It is quite possible that the gravitational field equations for
the spacetime metric have a predisposition to thermodynamic
behavior. This profound connection between gravity and
thermodynamics was first addressed by Jacobson who disclosed that
the hyperbolic second order partial differential Einstein equation
can be derived from the relation between the horizon area and
entropy, together with the Clausius relation $\delta Q=T\delta S$
\cite{Jac}. The investigations on the deep connection between
gravity and thermodynamics have been generalized to the
cosmological context where it has been shown that the differential
form of the Friedmann equation in the FRW universe can be written
in the form of the first law of thermodynamics on the apparent
horizon \cite{Cai2,Cai3,CaiKim,Fro,verlinde,Cai4,Shey1,Shey2}. See
\cite{Pad} for a recent review on thermodynamical aspects of
gravity.

If thermodynamical interpretation of gravity near the apparent
horizon is a generic feature, one needs to verify whether the
results may hold not only for more general spacetimes but also for
the other principles of thermodynamics, especially the generalized
second law (GSL) of thermodynamics as a global accepted principle
in the universe. The GSL of thermodynamics is an important
principle in governing the nature. Recently the GSL in the
accelerating universe enveloped by the apparent horizon has been
investigated in \cite{wang1,wang2,Shey33}. It was argued
\cite{wang2} that in contrast to the case of the apparent horizon,
the GSL of thermodynamics breakdown if one consider the universe
to be enveloped by the event horizon with the usual definitions of
entropy and temperature. This study reveals that in an
accelerating universe with spatial curvature, the apparent horizon
is a physical boundary from the thermodynamical point of view.
Using the general expression of temperature at apparent horizon of
FRW universe, it has been shown that the GSL holds in Einstein,
Gauss-Bonnet and more general lovelock gravity \cite{akbar}. The
GSL of thermodynamics has also been studied in the framework of
braneworld \cite{Shey3}. Other studies on the GSL of
thermodynamics have been carried out in
\cite{Pavon2,Karami,Other}.

It is interesting to note that Friedmann equations, in Einstein's
gravity, can be derived by using Clausius relation to the apparent
horizon of FRW universe, in which entropy is assumed to be
proportional to its horizon area, $S={A}/{4}$ \cite{CaiKim}.
However, this definition for entropy can be modified from the
inclusion of quantum effects, motivated from the loop quantum
gravity (LQG). The quantum corrections provided to the
entropy-area relationship leads to the curvature correction in the
Einstein-Hilbert action and vice versa \cite{Zhu}. The corrected
entropy takes the form \cite{Zhang}
\begin{equation}\label{ec}
S_A=\frac{A}{4}-\alpha \ln \frac{A}{4}+\beta \frac{4}{A},
\end{equation}
where $\alpha$ and $\beta$ are positive dimensionless constants of
order unity. The exact values of these constants are not yet
determined and still an open issue in loop quantum cosmology.
These corrections arise in the black hole entropy in LQG due to
thermal equilibrium fluctuations and quantum fluctuations
\cite{Rovelli}. Taking into account the quantum corrections in
entropy expression, it has been shown that the Friedmann equations
will be modified as well \cite{Cai5,Shey4,Shey5}. The cosmological
implications of the modified Friedmann equations have also been
studied in different setup \cite{Odintsov1,Odintsov2}.

In the present work we would like to examine whether the corrected
entropy-area relation, from LQG, together with the matter field
entropy inside the apparent horizon will satisfy the GSL of
thermodynamics. To be more general we will consider an interacting
viscous dark energy (DE) with dark matter (DM). In addition, we
will also include the contribution from the radiation. In an
isotropic and homogeneous FRW universe, the dissipative effects
arise due to the presence of bulk viscosity in cosmic fluids. The
theory of bulk viscosity was initially investigated by Eckart
\cite{Eck1} and later on pursued by Landau and Lifshitz
\cite{Lan}. DE with bulk viscosity has a peculiar property to
cause accelerated expansion of phantom type in the late evolution
of the universe \cite{Brevik}. It can also alleviate several
cosmological puzzles like age problem, coincidence problem  and
phantom crossing.

The structure of this paper is as follows. In the next section we
consider the modified Friedmann equation for a universe filled
with viscous DE, DM and radiation. In section 3, we examine the
validity of the GSL for the modified Friedmann equation
corresponding to the corrected entropy-area relation. We finish
with conclusions in the last section.

\section{Interacting viscous DE, DM and radiation in modified FRW cosmology}\label{DE}
In the framework of Friedmann–-Robertson–-Walker (FRW) metric,
\begin{equation}
{\rm d}s^2=-{\rm d}t^2+a^2(t)\left(\frac{{\rm d}r^2}{1-kr^2}+r^2{\rm
d}\Omega^2\right),\label{metric}
\end{equation}
for the non-flat FRW universe containing the DE, DM and radiation,
the modified first Friedmann equation corresponding to the
corrected entropy-area relation (\ref{ec}) is given by \cite{Cai5}
\begin{equation}
H^2+\frac{k}{a^2}-\frac{\alpha
}{2\pi}\left(H^2+\frac{k}{a^2}\right)^2- \frac{\beta
}{3\pi^2}\left(H^2+\frac{k}{a^2}\right)^3=\frac{8\pi
}{3}(\rho_D+\rho_m+\rho_r),\label{eqf1}
\end{equation}
where we take $G=1$ and $k=0,1,-1$ represent a flat, closed and
open universe, respectively. Also $\rho_D$, $\rho_m$ and $\rho_r$
are the energy density of DE, DM and radiation, respectively.

From Eq. (\ref{eqf1}), we can write
\begin{equation}
1+\Omega_k=\Omega_D+\Omega_m+\Omega_r+\Omega_\alpha+\Omega_\beta,
\label{eqf2}
\end{equation}
where we have used the following definitions
\begin{equation}
\Omega_{k}=\frac{k}{a^2H^2},~~~\Omega_{D}=\frac{8\pi
\rho_{D}}{3H^2},~~~\Omega_{m}=\frac{8\pi
\rho_{m}}{3H^2},~~~\Omega_{r}=\frac{8\pi
\rho_{r}}{3H^2},\label{Omega1}
\end{equation}
\begin{equation}
\Omega_\alpha=\frac{\alpha
}{2\pi}H^2(1+\Omega_k)^2,~~~\Omega_\beta=\frac{\beta
}{3\pi^2}H^4(1+\Omega_k)^3.\label{Omega2}
\end{equation}
Here we would like to consider the viscous DE model. Dissipative
processes are thought to be present in any realistic theory of the
evolution of the universe. The observations also indicate that the
universe media is not a perfect fluid and the viscosity is
concerned in the evolution of the universe (see \cite{Ren1} and
references therein). In the framework of FRW metric, the shear
viscosity has no contribution in the energy-momentum tensor, and
the bulk viscosity behaves like an effective pressure \cite{Ren2}.
The total energy density satisfies a conservation law
\begin{equation}
\dot{\rho}+3H(\rho+p)=0,\label{ed}
\end{equation}
where
\begin{equation}
\rho=\rho_D+\rho_m+\rho_r,\label{rho}
\end{equation}
\begin{equation}
p=\tilde{p}_D+p_r,\label{p1}
\end{equation}
and
\begin{equation}
\tilde{p}_D=p_D-3H\xi,
\end{equation}
is the effective pressure of the DE and $\xi$ is the bulk
viscosity coefficient \cite{Ren1,Ren2}. Note that the DM is
pressureless, i.e. $p_m=0$. Here like \cite{Sheykhi1}, if we
assume $\xi=\varepsilon\rho_D H^{-1}$, where $\varepsilon$ is a
constant parameter, then the total pressure can be written as
\begin{equation}
p=(\omega_D-3\varepsilon)\rho_D+\frac{1}{3}\rho_r,\label{p2}
\end{equation}
where $\omega_D=p_D/\rho_D$ is the equation of state (EoS)
parameter of the viscous DE.

We consider the case where the viscous DE, DM and radiation interact
with each other. In some recent studies, the scenario in which the
DE interacts with DM and radiation has been introduced to resolve
the cosmic triple coincidence problem \cite{triple}. Interaction
causes the DE, DM and radiation do not conserve separately and they
must rather enter the energy balances
\begin{equation}
\dot{\rho_D}+3H\rho_D(1+\omega_D)=9H^2\xi-Q',\label{continD}
\end{equation}
\begin{equation}
\dot{\rho_m}+3H\rho_m=Q,\label{continm}
\end{equation}
\begin{equation}
\dot{\rho_r}+4H\rho_r=Q'-Q,\label{continr}
\end{equation}
where $Q$ and $Q'$ stand for the interaction terms.

Taking time derivative in both sides of Eq. (\ref{eqf1}), and
using Eqs. (\ref{eqf2}), (\ref{Omega1}), (\ref{Omega2}),
(\ref{continD}), (\ref{continm}) and (\ref{continr}), one can get
the EoS parameter of interacting viscous DE as
\begin{eqnarray}
\omega_D=-\frac{1}{3\Omega_D}\left[2\left(\frac{\dot{H}}{H^2}-\Omega_k\right)\left(1-\frac{2\Omega_{\alpha}+3\Omega_{\beta}}{1+\Omega_k}\right)
+3\Omega_m+4\Omega_r\right]+3\varepsilon-1.\label{wlambda}
\end{eqnarray}
The deceleration parameter is given by
\begin{equation}
q=-\left(1+\frac{\dot{H}}{H^2}\right).\label{q1}
\end{equation}
Replacing the term $\dot{H}/H^2$ from (\ref{wlambda}) into
(\ref{q1}) yields
\begin{equation}
q=\frac{(1+\Omega_k)}{2(1+\Omega_k-2\Omega_{\alpha}-3\Omega_{\beta})}\Big[3\Omega_D(1+\omega_D-3\varepsilon)
+3\Omega_m+4\Omega_r\Big]-(1+\Omega_k).\label{q2}
\end{equation}
Using Eq. (\ref{eqf2}) one can rewrite (\ref{q2}) as
\begin{equation}
q=\frac{(1+\Omega_k)}{2(1+\Omega_k-2\Omega_{\alpha}-3\Omega_{\beta})}\Big[1+\Omega_{k}+\Omega_{\alpha}+3\Omega_{\beta}+3\Omega_D
(\omega_D-3\varepsilon)+\Omega_r\Big].\label{q3}
\end{equation}

\section{GSL with corrected entropy-area relation}\label{GSL}
Here, we study the validity of the GSL of thermodynamics for the
entropy-corrected Friedman equation. According to the GSL, entropy
of the viscous DE, DM and radiation inside the horizon plus the
entropy associated with the apparent horizon do not decrease with
time. Define $\tilde{r}(t) = a(t)r$, the metric (\ref{metric}) can
be rewritten as ${\rm d}s^2 = h_{ij}{\rm d}x^i{\rm d}x^j +
\tilde{r}^2{\rm d}\Omega^2$, where $x^i = (t, r)$, $h_{ij}$ =
diag($-1, a^2/(1 - kr^2)$), $i,j=0,1$. By definition
\begin{equation}
f:=h^{ij}\partial_{i}\tilde{r}\partial_{j}\tilde{r}=1-\left(H^2+\frac{k}{a^2}\right)\tilde{r}^2,\label{f}
\end{equation}
when $f=0$ then the location of the apparent horizon in the FRW
universe is obtained as \cite{Poisson}
\begin{equation}
\tilde{r}_{\rm A}=H^{-1}(1+\Omega_k)^{-1/2}.\label{ra}
\end{equation}
The surface gravity for the FRW universe is obtained as
\cite{CaiKim}
\begin{equation}
\kappa=-\frac{1}{\tilde{r}_{\rm
A}}\left(1-\frac{\dot{\tilde{r}}_{\rm A}}{2H\tilde{r}_{\rm A}}
\right).
\end{equation}
The associated Hawking temperature on the apparent horizon is
defined as
\begin{equation}
T_{\rm A}=\frac{|\kappa|}{2\pi}=\frac{1}{2\pi \tilde{r}_{\rm
A}}\left(1-\frac{\dot{\tilde{r}}_{\rm A}}{2H\tilde{r}_{\rm A}}
\right),\label{TA1}
\end{equation}
where $\frac{\dot{\tilde{r}}_{A}}{2H\tilde{r}_{A}}<1$ ensure that
the temperature is positive.

Taking time derivative in both sides of Eq. (\ref{ra}), and using
Eqs. (\ref{eqf1}), (\ref{eqf2}), (\ref{Omega1}), (\ref{Omega2}),
(\ref{ed}), (\ref{rho}) and (\ref{p2}), one can get
\begin{equation}
{\dot{\tilde{r}}_A}=\frac{(1+\Omega_k)^{-1/2}}{2(1+\Omega_k-2\Omega_{\alpha}-3\Omega_{\beta})}\Big[3\Omega_D(1+\omega_D-3\varepsilon)
+3\Omega_m+4\Omega_r\Big].\label{radot1}
\end{equation}
Using Eq. (\ref{q2}) one can rewrite (\ref{radot1}) as
\begin{equation}
{\dot{\tilde{r}}_A}=\frac{1+\Omega_k+q}{(1+\Omega_k)^{3/2}}.\label{radot2}
\end{equation}
From Eqs. (\ref{ec}), (\ref{TA1}) and using $A=4\pi
\tilde{r}_A^2$, the evolution of the apparent horizon entropy is
obtained as
\begin{equation}
T_A\dot{S}_A=4\pi
H{\tilde{r}}_A^3(\rho+p)-2\pi{\tilde{r}}_A^2{\dot{\tilde{r}}}_A(\rho+p).\label{em}
\end{equation}
The entropy of the universe including the viscous DE, DM and
radiation inside the dynamical apparent horizon can be related to
its energy and pressure in the horizon by Gibbs equation
\cite{Pavon2}
\begin{equation}
T{{\rm d}S}={\rm d}(\rho V)+p~{\rm d}V=V{\rm d}\rho+(\rho+p){\rm
d}V,\label{Gibbs}
\end{equation}
where ${\rm V}=4\pi \tilde{r}_{\rm A}^3/3$ is the volume of the
universe enclosed by the dynamical apparent horizon
$\tilde{r}_{\rm A}$. Following \cite{Shey3,Karami}, we limit
ourselves to the assumption that the thermal system bounded by the
dynamical apparent horizon remain in equilibrium so that the
temperature of the system must be uniform and the same as the
temperature of its boundary. This requires that the temperature
$T$ of the universe enclosed by the dynamical apparent horizon
should be in equilibrium with the Hawking temperature $T_{\rm A}$
associated with the dynamical apparent horizon, so we have $T =
T_{\rm A}$. Therefore from Eq. (\ref{Gibbs}) one can obtain
\begin{equation}
T_A\dot{S}=4\pi{\tilde{r}}_A^2{\dot{\tilde{r}}_A}(\rho+p) -4\pi
H{\tilde{r}}_A^3(\rho+p),\label{en}
\end{equation}
where $S=S_D+S_m+S_r$ is the entropy in the universe containing
the viscous DE, DM and radiation. Finally, adding  Eqs. (\ref{em})
and (\ref{en}), the GSL due to the different contributions of the
viscous DE, DM, radiation and dynamical apparent horizon can be
obtained as
\begin{equation}
T_A\dot{S}_{\rm
{tot}}=2\pi{\tilde{r}}_A^2{\dot{\tilde{r}}_A}(\rho+p),\label{eu}
\end{equation}
where $S_{\rm tot}=S+S_{\rm A}$ is the total entropy.

From Eqs. (\ref{rho}), (\ref{p2}) and using (\ref{Omega1}) one can
obtain
\begin{equation}
\rho+p=\frac{H^2}{8\pi}\Big[3\Omega_D(1+\omega_D-3\varepsilon)
+3\Omega_m+4\Omega_r\Big].\label{rhop1}
\end{equation}
Using Eq. (\ref{q2}) one can rewrite (\ref{rhop1}) as
\begin{equation}
\rho+p=\frac{H^2}{4\pi
}\left(\frac{1+\Omega_k+q}{1+\Omega_k}\right)(1+\Omega_k-2\Omega_{\alpha}-3\Omega_{\beta}).\label{rhop2}
\end{equation}
Substituting Eqs. (\ref{ra}), (\ref{radot2}) and (\ref{rhop2})
into (\ref{eu}) yields the GSL as
\begin{equation}
T_A\dot{S}_{\rm
{tot}}=\frac{(1+\Omega_k-2\Omega_\alpha-3\Omega_\beta)}{2(1+\Omega_{k})^{7/2}}(1+\Omega_k+q)^2,\label{GSL3}
\end{equation}
which can be rewritten by the help of Eq. (\ref{q2}) as
\begin{equation}
T_A\dot{S}_{\rm
{tot}}=\frac{1}{8}\frac{\Big[3\Omega_D(1+\omega_D-3\varepsilon)
+3\Omega_m+4\Omega_r\Big]^2}{(1+\Omega_{k})^{3/2}(1+\Omega_k-2\Omega_\alpha-3\Omega_\beta)}.\label{GSL4}
\end{equation}
According to Eq. (\ref{GSL4}), the validity of GSL, i.e.
$T_A\dot{S}_{\rm {tot}}>0$, depends on the sign of expression
$(1+\Omega_k-2\Omega_\alpha-3\Omega_\beta)$ appeared in the
denominator. From Eqs. (\ref{Omega2}) and (\ref{ra}) we have
\begin{equation}
1+\Omega_k-2\Omega_\alpha-3\Omega_\beta=(1+\Omega_k)\left(1-\frac{\alpha
}{\pi{\tilde{r}}_A^2}-\frac{\beta
}{\pi^2{\tilde{r}}_A^4}\right).\label{cond1}
\end{equation}
Following \cite{Shey5}, it has been shown that
\begin{equation}
\left(1-\frac{\alpha}{\pi{\tilde{r}}_A^2}-\frac{\beta
}{\pi^2{\tilde{r}}_A^4}\right)>0,\label{cond2}
\end{equation}
because $\tilde{r}_A\gg 1$ while $\alpha\sim O(1)$ and $\beta\sim
O(1)$, thus $\frac{\alpha}{\pi{\tilde{r}}_A^2}\ll 1$ and
$\frac{\beta}{\pi^2{\tilde{r}}_A^4}\ll 1$.

Finally from Eqs. (\ref{GSL4})-(\ref{cond2}) one can conclude that
$T_A\dot{S}_{\rm {tot}}>0$. Therefore the GSL for the modified FRW
universe containing the interacting viscous DE with DM and
radiation enclosed by the dynamical apparent horizon is always
satisfied throughout the history of the universe for any spatial
curvature and it is independent of the EoS parameter of
interacting viscous DE model.

\section{Conclusions}
Here, we investigated the validity of the GSL of thermodynamics
for an entropy-corrected FRW universe filled with viscous DE and
DM as well as radiation. Following the method developed in
\cite{Shey3}, we examined time evolution of the total entropy
including the entropy associated with the apparent horizon plus
the entropy of viscous DE, DM and radiation inside the apparent
horizon. We found out that the GSL of thermodynamics in a modified
FRW universe with entropy correction terms is fulfilled throughout
the history of the universe. The satisfaction of the GSL of
thermodynamics for the modified FRW cosmology further supports the
thermodynamical interpretation of gravity and provides more
confidence on the profound connection between gravity and
thermodynamics.
\\
\\
\noindent{\textbf{Acknowledgements}. The works of K. Karami and A.
Sheykhi have been supported financially by Research Institute for
Astronomy and Astrophysics of Maragha, Iran.


\end{document}